\documentclass[aps,preprint,prd,nofootinbib]{revtex4-1}
\usepackage{latexsym,bm,amsmath,amssymb,xfrac,xcolor,graphicx,epstopdf}
\usepackage[linktoc=all,hidelinks]{hyperref}
\usepackage{multirow}
\DeclareMathAlphabet{\mathpzc}{OT1}{pzc}{m}{it}

\newcommand*\dif{\mathop{}\!\mathrm{d}}
\newcommand{\link}[1]{[\href{http://arxiv.org/abs/#1}{{\tt arXiv:#1}}]}
\newcommand{\linkth}[1]{[\href{http://arxiv.org/abs/hep-th/#1}{{\tt arXiv/hep-th:#1}}]}

\usepackage[shortlabels]{enumitem}

\def\p{\partial}
\def\m{\mu}
\def\n{\nu}
\def\nab{\nabla}
\def\a{\alpha}
\def\b{\beta}

\def\el{\ell}
\def\ep{\epsilon}

\def\L{\Lambda}

\def\ps{\psi}
\def\r{\rho}
\def\s{\sigma}

\def\t{\theta}

\def\cO{\mathcal{O}}

\def\goesto{\rightarrow}

\usepackage{geometry}
\begin{document}
		\title{Brief Note on Thurston Geometries in 3D Quadratic Curvature Theories}
	\author{G{\"o}khan Alka\c{c}}
	\email{gokhanalkac@hacettepe.edu.tr}
	
	\affiliation{Physics Engineering Department, Faculty of Engineering, Hacettepe
		University, 06800, Ankara, Turkey}
	\author{Deniz Olgu Devecio\u{g}lu}
	 \email{dodeve@gmail.com}
	
	\affiliation{School of Physics, Huazhong University of Science and Technology,\\
		Wuhan, Hubei,  430074, China }
	\date{\today}
	
\begin{abstract}
We show that Thurston geometries are solutions to a large class of 3D quadratic curvature theories, where New Massive Gravity, which was studied in \cite{Flores-Alfonso:2021opl}, is a special case. 
\end{abstract}
	\maketitle	

Thurston's conjecture \cite{Thurston} states that a three-manifold with a
given topology has a canonical decomposition into eight manifolds with high symmetry, the so-called Thurston geometries (TGs) \cite{Thurston2,Thurston3,Thurston4}. Based on Hamilton's Ricci flow \cite{Hamilton}, which provides a way to deform a given metric in the form of a generalized heat equation, Perelman gave his celebrated proof \cite{Perelman1,Perelman2}. Since many of TGs are not Einstein spaces, one might wonder if a `master theory' that admits all TGs as solutions can be achieved by a modification of Einstein's equation.

In a recent letter \cite{Flores-Alfonso:2021opl}, it was reported that TGs are solutions to New Massive Gravity (NMG) \cite{Bergshoeff:2009hq}, where the authors state that, to the best of their knowledge, NMG is the only 3D gravity theory with this property\footnote{In \cite{Flores-Alfonso:2021opl}, ghost-free theories were considered for simplicity  \cite{Flores-AlfonsoP}.}. In a previous work \cite{Flores-Alfonso:2020zif}, four of the TGs were shown to be
solutions of the most general quadratic theory. In this paper we shall
close a gap and show that the remaining four TGs are also solutions of
the most general quadratic theory, whose action is displayed in \eqref{action}. NMG is singled out by being the only ghost-free theory in this class\footnote{NMG is obtained after imposing  the absence of the scalar mode, which is always a ghost, and the unitarity of the massive spin-2 mode around a maximally symmetric background spacetime (see \cite{Gullu:2010sd} for an analysis on a flat background and \cite{Bergshoeff:2009aq,Tekin:2016vli} for constant curvature backgrounds (dS/AdS)).}. However, ghost-freedom is not essential in defining a flow where TGs are critical points.

In order to demonstrate this, let us present the logic of a uniformization theorem along the lines of \cite{Gegenberg:2003yz}. Assuming that the manifold admits some metric, one needs to devise a mechanism to flow the initial metric into a highly symmetric one, which can be obtained by the following parabolic system
\begin{equation}
\p_t g_{\m\n} = \cO(g_{\m\n}),\label{flow}
\end{equation}
where $\cO$ is an elliptic operator. The parabolicity of \eqref{flow} assures that the flow will end at fixed points as $t \goesto \infty$, for which $\cO(g_{\m\n})=0$. However, since $\cO(g_{\m\n})$ should transform as a tensor, it has to be constructed from curvature tensors, which makes it impossible to obtain a strictly parabolic system and leads to nonlinearity due to the inverse metric. Many issues regarding these complications should be resolved for a full proof. In his proof, Perelman made use of the operator
\begin{equation}
	\cO(g_{\m\n})=-2\left(R_{\m\n} +\nab_{\m}\nab_{\n}f\right),
\end{equation}
 where $f$ is a scalar field. It follows from the field equations of the action
 \begin{equation}
 	S[f,g] = \int \dif^3 x \sqrt{|g|}\, e^{-f}\left[R+(\nab f)^2\right],\label{actionp}
 \end{equation}
and the action \eqref{actionp} is an increasing functional provided that $\partial_t f = \partial_t \ln \sqrt{|g|}$. Alternatively, one can also try to use the field equations of quadratic gravity theories as a candidate for such an operator, which can be defined through an action that is a functional of the metric tensor only. As long as one finds an operator that vanishes for all TGs, the ghost-freedom of the theory plays no role in the construction.

Motivated by this, we consider the most general 3D quadratic gravity theory described by the action
\begin{equation}
S[g] = \int \dif^3 x \sqrt{|g|} \left[\sigma (R - 2\Lambda)+ \alpha R^2 + \beta R_{\mu\nu}R^{\mu\nu}\right]\,,\label{action}
\end{equation}
where $\sigma=\pm1$ is introduced to control the sign of the Einstein-Hilbert term and $(\alpha,\beta)$ are arbitrary constants. Field equations arising from the action \eqref{action} are given by
\begin{align}
	0  &=\s \left(G_{\mu\nu} + \Lambda g_{\mu\nu}\right) + \alpha \left[2 RR_{\mu\nu} - \dfrac12 g_{\mu\nu}R^2 + 2g_{\mu\nu} \square R - 2 \nabla_{\mu} \nabla_{\nu} R \right] \\ \nonumber
	&+ \beta \left[\dfrac32 g_{\mu\nu} R_{\rho\sigma} R^{\rho\sigma} 
	-4R_{\mu}^{\ \rho} R_{\nu\rho} + \square R_{\mu\nu} + \dfrac12 g_{\mu\nu}\square R - \nabla_{\mu}\nabla_{\nu}R + 3 RR_{\mu\nu} - g_{\mu\nu}R^2\right].
\end{align}
As we summarize in Tables I and II, TGs are solutions for different choices of ($\s,\alpha,\beta$) that are subjected to two constraints:
\begin{enumerate}[i.] 
	\item The length scale in the metric should satisfy $\el^2>0$ in order to consider metrics with definite signature. 
	\item Denominators in the expressions for $\el^2$ and $\L$ should be non-zero.
\end{enumerate}
As a simple example, to ensure that the theory admits Lorentzian TGs given in Table II as solutions, the following choice of parameters is enough
\begin{equation}
	\s = -1, \qquad \b = \dfrac{1}{m^2} >0, \qquad \a > -\dfrac{1}{2 m^2}.\label{constex}
\end{equation}

When one demands to obtain Lorentzian TGs as solutions to a pure gravity theory (no scalar mode) with unitary excitations around a maximally symmetric spacetime, then NMG is the only choice and one has to fix the last parameter as $\a=-\dfrac{3}{8m^2}$ \cite{Bergshoeff:2009hq,Bergshoeff:2009aq,Tekin:2016vli}, for which the scalar mode decouples from the spectrum since it becomes infinitely heavy ($m_s^2 \goesto \infty$), and there remains only a unitary massive spin-2 graviton with $m_g^2 = m^2 + \dfrac{\L}{2}>0$. 

For the Euclidean TGs, some possible choices of parameters are given in Table III, where NMG is again a special case. Note that $m^2<0$ is just having an imaginary mass. Analogous behaviour was also observed in \cite{Gegenberg:2002xj,Gegenberg:2003yz}, where the Euclidean solutions are obtained from the low energy limit of string theory supplemented by a one-form gauge field. For some of the solutions, gauge fields become imaginary, which is equivalent to having the kinetic terms with opposite signs in the action. However, this should not be considered as a problem since the aim here is to choose the parameters according to Table I and show that TGs correspond to fixed points.

TGs were also considered in another modification of Einstein's equations called Minimal Massive Gravity (MMG) \cite{Bergshoeff:2014pca}, whose field equations read
\begin{equation}
G_{\m\n} + \L\,g_{\m\n}+ \frac{1}{\m} C_{\m\n} + \frac{\gamma}{\m^2} J_{\m\n}=0, \label{MMG}
\end{equation}
where the Cotton tensor $C_{\m\n}$ is related to the Schouten tensor $S_{\m\n}$ as
\begin{equation}
	C_{\m\n}= \ep_{\m}^{\ \a \b} \nab_{\a} S_{\b\n}, \qquad S_{\m\n}=R_{\m\n}-\frac{1}{4} g_{\m\n} R,
\end{equation}
and the J-tensor is given by 
\begin{equation}
	J_{\m\n}= \epsilon_{\m}^{\ \a\b} \epsilon_{\n}^{\ \r\s} S_{\a\r} S_{\b\s}.
\end{equation}
The field equations \eqref{MMG} do not arise from the variation of an action and are covariantly conserved only on-shell. In \cite{Charyyev:2017uuu}, it was shown that all TGs except Sol geometry are solutions to \eqref{MMG} for certain (non-zero and finite) choice of parameters. Sol geometry requires the vanishing of the coefficient of the Cotton term ($\m \goesto \infty$, $\gamma \goesto \infty$ such that $\frac{\gamma}{\mu^2} \goesto \text{constant}$), which in general ruins the physical consistency of the theory. However, such a solution is still acceptable if $\epsilon^{\mu \rho \sigma} S_{\rho}^{\ \tau} C_{\sigma \tau}=0$, which is the case for the Sol geometry. Therefore, in addition to NMG, MMG defines an operator $\cO(g_{\m\n})$ that vanishes for all TGs. What field equations of MMG lacks is the existence of an entropy functional along the flow which admits a gradient formulation (see \cite{Kisisel:2008jx} for an example in Cotton flow), which might be crucial in an attempt for a full proof of a uniformization theorem. 

All in all, realising TGs as vacuum solutions to higher curvature modifications of Einstein's equations yield many possibilities where the ghost-free theories are not necessarily privileged. Due to the higher-derivative operators, it seems impossible to apply results from elliptic operator theory for rigorous results. However, the relevance of the unitarity of the theory might be checked through a detailed numerical analysis of the flow equation as done in \cite{Kisisel:2008jx} for the Cotton flow.

	\begin{table}[]
		\begin{center}
			\scalebox{0.9}{
		\begin{tabular}{|c|c|c|c|}
			\hline
			Geometry	& Metric &$\el^2$  & $\L$  \\\hline
			$\text{E}^3$  & $\dif s^2 = \dif x^2 + \dif y^2+ \dif z^2$ & no length scale  & 0  \\ \hline
			$\text{S}^3$	& $\dif s^2 = \el^2 \left(\dif x^2 + \sin^2 x \dif y^2 + \sin^2 x \sin^2 y \dif z^2\right)$          & arbitrary  & $\dfrac{1}{\el^2} -\s \dfrac{2(3\a + \b)}{\el^4}$  \\ \hline
			$\text{H}^3$	& $\dif s^2 = \el^2 \left(\dif x^2 + \cosh^2 x  \dif y^2 + \cosh^2 x  \cosh^2 y \dif z^2\right)$           & arbitrary  & $-\dfrac{1}{\el^2} -\s \dfrac{2(3\a + \b)}{\el^4}$  \\ \hline
			$\text{E}^1 \times \text{S}^2$	& $\dif s^2 = \el^2 \left(\dif x^2 +  \dif y^2 + \sin^2 y \dif z^2\right)$          & $\dfrac{1}{2\L}$  & $-\dfrac{\s}{4(2\a+\b)}$  \\ \hline
			$\text{E}^1 \times \text{H}^2$	&   $\dif s^2 = \el^2 \left(\dif x^2 +  \dif y^2 + \cosh^2 y \dif z^2\right)$        & $-\dfrac{1}{2\L}$  & $-\dfrac{\s}{4(2\a+\b)}$ \\ \hline
			Nil	&  $\dif s^2 = \dfrac{\el^2}{4} \left[\dif x^2 + \dif y^2 + (x \dif y - \dif z)^2\right]$& $-\dfrac{1}{2 \L}$ & $-\dfrac{\s}{8(\a+3\b)}$  \\ \hline
			SL(2,$\mathbb{R}$)	& $\dif s^2 = \el^2 \left[\dif r^2 + \sinh^2 r  \cosh^2 r  \dif \t^2+(\dif \ps+\sinh^2 r \dif \t)^2\right]$   & $-\dfrac{25\a+23\b}{10\L (\a+\b)}$  & $-\s \dfrac{25\a+23\b}{200(\a+\b)^2}$ \\ \hline
			Sol	& $\dif s^2 = \el^2 \left(e^{-2z} \dif x^2 + e^{2z} \dif y^2+\dif z^2 \right)$         & $-\dfrac{1}{2\L}$   & $-\dfrac{\s}{8(\a+\b)}$  \\ \hline
		\end{tabular}}\caption{Euclidean Thurston Geometries as solutions to an arbitrary quadratic gravity theory. Parameters should satisfy the constraints (i,ii), where examples of sufficient but not necessary conditions are given in Table III.}
	\end{center}
\end{table}

\begin{table}[]
	\begin{center}
		\scalebox{0.84}{
			\begin{tabular}{|c|c|c|c|}
				\hline
				Geometry	& Metric &$\el^2$  & $\L$  \\\hline
				Nil	&  $\dif s^2 = \dfrac{\el^2}{4} \left[\dif x^2 + \dif y^2 - (x \dif y - \dif z)^2\right]$& $\dfrac{1}{2 \L}$ & $-\dfrac{\s}{8(\a+3\b)}$  \\ \hline
				SL(2,$\mathbb{R}$)	& $\dif s^2 = \dfrac{\el^2}{4} \left[-(\dif \Psi +\cos \Theta \dif \Phi)^2+\dif \Theta^2 + \sin^2 \Theta \dif \Phi^2\right]$   & $\dfrac{25\a+23\b}{10\L (\a+\b)}$  & $-\s \dfrac{25\a+23\b}{200(\a+\b)^2}$ \\ \hline
				New Sol	& $\dif s^2 = \el^2 \left(e^{-2z} \dif x^2 + e^{2z} \dif y^2-\dif z^2 \right)$         & $\dfrac{1}{2\L}$   & $-\dfrac{\s}{8(\a+\b)}$  \\ \hline
				Lorentz Sol & $\dif s^2 = \el^2 \left(2 e^{-z}\dif x \dif z+e^{2z} \dif y^2\right) $  & arbitrary & $\s=0$, no effect  on the solution  \\ \hline
				Third Sol & $\dif s^2 = \el^2 \left(-e^{2z} \dif y^2 -2 \dif x \dif y + \dif z^2\right)$ & $-4 \s \b$ & 0 \\ \hline
				Lorentz-Heisenberg &  $\dif s^2 = \dfrac{\el^2}{4} \left(-\dif x^2 + \dif y^2 + (x \dif y - \dif z)^2\right)$& $\dfrac{1}{2 \L}$ & $-\dfrac{\s}{8(\a+3\b)}$    \\ \hline
				AdS & $\dif s^2  = \el^2 \left[\dif r^2+ \sinh^2 r \cosh^2 r \dif \t^2-(\dif \ps + \sinh^2 r \dif \t)^2\right]$  & arbitrary  & $-\dfrac{1}{\el^2} -\s \dfrac{2(3\a + \b)}{\el^4}$   \\ \hline
		\end{tabular}}\caption{Lorentzian Thurston Geometries as solutions to an arbitrary quadratic gravity theory. Parameters should satisfy the constraints (i,ii), where an example of sufficient but not necessary conditions is given in \eqref{constex}.}
	\end{center}
\end{table}

\begin{table}[]
	\begin{tabular}{|c|c|}
		\hline
	Geometry	& Parameters  \\ \hline
		\multirow{2}{*}{$\text{E}^1 \times \text{S}^2$} & $\qquad\s=-1\qquad$ $\b=\dfrac{1}{m^2}>0\qquad$ $\a>-\dfrac{1}{2m^2}\qquad$  \\ \cline{2-2} 
		& $\s=+1\qquad$ $\b=\dfrac{1}{m^2}<0\qquad$ $\,\a<-\dfrac{1}{2m^2}$  \\ \hline
		\multirow{2}{*}{Others} & $\qquad\s=-1\qquad$ $\b=\dfrac{1}{m^2}<0\qquad$ $\a<-\dfrac{1}{2m^2}\qquad$  \\ \cline{2-2} 
		& $\qquad\s=+1\qquad$ $\b=\dfrac{1}{m^2}>0\qquad$ $\a>-\dfrac{1}{2m^2}\qquad$ \\ \hline
	\end{tabular}\caption{Choice of parameters for Euclidean Thurston Geometries. Note that these are sufficient but not necessary conditions.}
\end{table}
\newpage
{\bf Acknowledgements: }D. O. D.  was supported in part by the National Natural Science Foundation of China under Grant No. 11875136 and the Major Program of the National Natural Science Foundation of China under Grant No. 11690021. We thank Nihat Sad{\i}k De\u{g}er and Daniel Flores-Alfonso for useful discussions. We thank the anonymous referee for his comments, which led us to improve the presentation of our work.

\end{document}